\documentclass{emulateapj}

\newcommand{\msun}{M_{\sun}}

\begin{document}

\title{Origin of the S Stars in the Galactic Center}

\author{Ulf L\"ockmann, Holger Baumgardt, and Pavel Kroupa}
\affil{Argelander-Institut f\"ur Astronomie, Universit\"at Bonn, Auf dem H\"ugel 71, 53121 Bonn, Germany}
\email{uloeck@astro.uni-bonn.de}

\begin{abstract}
Over the last 15 years, around a hundred very young stars have been observed in the central parsec of our Galaxy.
While the presence of young stars forming one or two stellar disks at $\approx 0.1\,$pc from the supermassive black 
hole (SMBH) can be understood through star formation in accretion disks,
the origin of the S stars observed a factor of 10 closer to the SMBH has remained a major puzzle.
Here we show the S stars to be a natural consequence of dynamical interaction of two stellar disks at larger radii.
Due to precession and Kozai interaction,
individual stars achieve extremely high eccentricities at random orientation.
Stellar binaries on such eccentric orbits are disrupted due to close passages near the SMBH, leaving behind a single S star on a much tighter orbit.
The remaining star may be ejected from the vicinity of the SMBH,
thus simultaneously providing an explanation for the observed hypervelocity stars in the Milky Way halo.
\end{abstract}


\section{Introduction}

The central parsec of our Galaxy harbors different groups of young stars orbiting the central SMBH.
These include one or two stellar disks at $\approx 0.1\,$pc from Sgr A$^*$ made up of 5--6\,Myr old stars \citep{pau06},
and a number of stars on highly eccentric orbits even closer, the so-called S stars \citep{eg97}.
With orbital periods as small as 15 years, the S stars provide powerful constraints on the mass and size of the central dark object \citep{ghe05}.
While the origin of the stellar disk(s) observed in the Galactic center can be understood \citep{lb03,gs03,nc05,ger01,mp03},
the S stars show a puzzling combination of interesting features \citep{eg97}: They are very young ($< 10$\,Myr), with less than 0.01\,pc a factor of 10 closer to the SMBH than the closest young population known (the stellar disks), and move around Sgr A$^*$ on randomly oriented eccentric orbits.

A number of mechanisms to create the S stars have been proposed, but so far none of them was able to account for all their properties \citep[see also review by][]{a05}:
It has been suggested that the S stars are not young but rejuvenated stars or exotic objects,
but this is
in conflict with their spectral properties, showing they are ordinary main-sequence B-type stars \citep{ghe05,eis05,mge08}.
In situ formation is excluded by the inexplicably high pressure required to overcome the strong tidal forces, and will also lead to a stellar disk \citep{ghe05},
which cannot account for the random orientation and high eccentricity of the orbits within the stars' lifetime \citep{gh07}.
The latter argument also argues against the suggested scenario of an infalling cluster of young stars \citep{ger01},
which would moreover require a very high mass \citep{pmg03}, or an intermediate-mass black hole in its center for it to survive the tidal force of Sgr A$^*$ \citep{hm03}.
Exchange capture with compact remnants \citep{al04} only works with a large cluster of black holes and fails to explain the youth and high initial eccentricities.
The Lense-Thirring effect may account for the innermost star \citep{lb03}, S2, being dragged out of the disk plane,
but cannot explain the orbital orientation of the longer-period S stars, nor the proximity of the orbit itself.
Interaction of stars in a central cluster with a fossil accretion disk can lead to high eccentricities and orbital decay \citep{sk05} if the disk mass is sufficiently high, but fails to explain the observed bias toward young B-type stars. Furthermore, such an accretion disk is expected to fragment and form stars and should thus have been observed.

Tidal disruption of binaries falling into the Galactic center on highly eccentric orbits has been shown to be an effective mechanism to create S stars,
but until now required very massive star clusters
passing close to Sgr A$^*$ \citep{gq03}
or the presence of a large enough population of compact massive perturbers in combination with many young stars on eccentric orbits \citep{pha06}.

In the following we show how the dynamical interaction of two stellar disks at larger radii, as they are observed in the Galactic center, naturally leads to the creation of S stars. Section \ref{sec:model} describes the setup of our numerical simulations. In \S\ \ref{sec:high-ecc}, we present the outcome of these simulations, showing how the disk stars can eventually gain very high eccentricities. Section \ref{sec:disrupt} explains how binary stars on such eccentric orbits around the SMBH get disrupted, leaving behind an S star and possibly a hypervelocity star. In \S\ \ref{sec:discuss} we discuss our results, showing how the mechanism suggested here explains all the observed properties of the S stars.

\section{Modeling disk interaction}
\label{sec:model}

\begin{figure}
\plotone{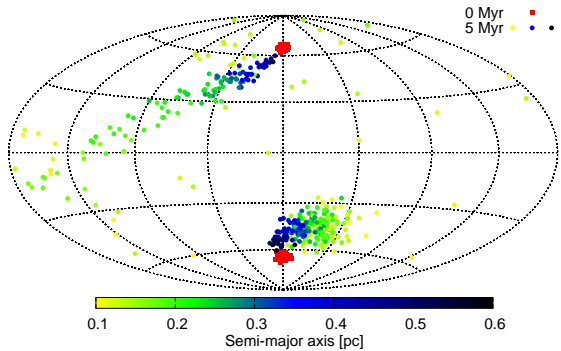}
\caption{
\label{fig:discevol1}
Aitoff projection of normal vectors of the orbits of stars situated in two interacting disks.
The separation of two points in the projected plot indicates the relative inclination of the two corresponding stellar orbits,
and thus the disk thickness.
The red squares indicate the initial state of two nearly flat disks inclined at an angle of 130$^{\circ}$.
The orientation of the disks changes with time due to precession.
By 5\,Myr (circles), the massive disk shows a slight warping (bottom).
The less massive disk (top to left) precesses faster and is affected more strongly by warping,
especially towards the inner edge (green and yellow circles), where it loses its disk shape.}
\end{figure}

An extensive analysis of young massive stars in the innermost parsec of our Galaxy shows that
almost all stars reside in two $6 \pm 2$\,Myr old rotating disks---except for the S stars in the central arcsecond \citep{pau06}.
While the clockwise disk is very distinct, there is some debate over the disk structure of the counterclockwise system \citep{lu06}.
However, \citet{caa08} have shown that a single stellar disk cannot explain the high inclinations and eccentricities observed.
Using direct $N$-body calculations as described below, we find that the distribution of eccentricities and inclinations of the observed young stars' orbits is consistent with an initial configuration of two relatively flat disks, since the less massive disk is destroyed over time due to dynamical interaction (see \S\ \ref{sec:high-ecc}).

To investigate the stellar dynamical evolution of two interacting stellar disks,
we have computed a number of $N$-body integrations. The parameters of our model are 
chosen so as to best fit the observations by \citet{pau06} and are as follows:
Our model consists of a $3.5 \times 10^6\,\msun$ SMBH and two flat circular stellar disks, mutually inclined at an angle of $130^{\circ}$, 
with a surface density profile that scales with distance as $r^{-2.5}$. The disks have well-defined initial extents
of $0.05-0.5\,$pc and $0.07-0.5\,$pc, respectively.

We assume that the lower mass limit of the stars observed by \citet{pau06} is $20\,\msun$. For our models, we assume a disk age of 5\,Myr, and 
hence an upper initial mass limit of $42.5\,\msun$ for the observed disk stars \citep{hpt00,fba06}. Using the initial mass function
${\rm d}N/{\rm d}m \propto m^{-1.35}$ as derived from the $K$-band luminosity function by \citet{pau06} with mass limits 1 and 
120\,$\msun$, this leads to a total initial stellar mass of $10,380\,\msun$ to explain the 73 observed O stars. Allowing for 
30\% binaries and another 10\% for stars disrupted by the SMBH or unobserved, and using a mass ratio of 2\,:\,1 for the disks, we have 
initial stellar masses of $9,900$ and $4,950\,\msun$, respectively. As a consequence, our disks initially consist of 535 and 267 multimass 
stars (and binaries), respectively.
The above parameters have proven to result in 
a distribution of eccentricities and semimajor axes as well as an inclination between the two disks after 5 Myr which best match
the observations.

We used our \textsc{bhint} code \citep{lb08} for the integration, which has been developed specifically to calculate the dynamics of 
stars orbiting a SMBH, and includes post-Newtonian treatment up to order 2.5 to account for the effects of general relativity.
Furthermore, we have added the \textsc{SSE} package by \citet{hpt00} to account for the effects of stellar mass loss.

The stars and binaries are modeled as point masses, and all binaries are assumed to be equal-mass.
Tidal disruption is not considered; however, we assume that stars on very tight orbits ($a<80$\,AU) are swallowed by the SMBH.
We switch on the post-Newtonian terms for the central motion around the SMBH in our \textsc{bhint} integrator \citep{lb08} 
for eccentricities $e>0.9$, and switch it off when $e$ falls below 0.8. A fully post-Newtonian calculation shows that in the regime 
considered in this Letter, relativistic effects are negligible for orbits with lower eccentricities within the considered timescale.

We finally note that the detailed choice of parameters seems not to be too important. A number of models with variations in total mass, 
mass function, and disk age lead to comparable results.

\section{Achieving very high eccentricities}
\label{sec:high-ecc}

Due to mutual torques, the two stellar disks described above precess about each other, with a frequency proportional to the other disk's mass:
The precession frequency of a star orbiting a SMBH of mass $M_{\rm SMBH}$ at radius $R$ at an inclination $\beta$ relative to a narrow disk of mass $M_{\rm disk}$ at a radius $R_{\rm disk}$ can be approximated as \citep{nay05}
\begin{equation}
\omega_p = -\frac{3}{4}\frac{M_{\rm disk}}{M_{\rm SMBH}}\cos{\beta}\sqrt{\frac{GM_{\rm SMBH}}{R^3}}\frac{R^3R_{\rm disk}^2}{\left(R^2+R_{\rm disk}^2\right)^{5/2}}.
\end{equation}

As can be seen, the
precession frequency depends also on the distance of a star from the central SMBH \citep{nay05},
and since both disks have a finite extent, stars at different central distances precess with different frequencies,
thus warping the originally flat disks. In particular, the innermost stars quickly get inclined with respect to their disk of origin,
and so start precessing about both disks. This leads to further thickening and deformation of the disks as well as large inclinations.

Figure \ref{fig:discevol1} depicts the evolution of the stellar disks' shape and thickness.
Being exposed to the larger torque, the less massive disk is warped much faster,
explaining why the observed thickness of the less massive disk
in the center of our Galaxy is substantially higher, making its present existence
controversial \citep{lu06}.

Apart from precession, all stars are subject to the Kozai mechanism \citep{k62}: Stars on highly inclined orbits relative to an axisymmetric perturbation periodically obtain small inclinations, and simultaneously gain high eccentricities.
At their pericenter, stars on orbits with eccentricity $e>0.99$ pass the central SMBH within $\sim 100$\,AU, less than the S stars' pericenter distances.
The effects of general relativity causes eccentric orbits to precess, which may damp the Kozai mechanism and prevent further growth in eccentricity \citep{htt97}.
However, our simulations show that in the presence of very massive, nearby disks, relativistic effects do not prevent eccentricities as high as $e=0.999$.

\begin{figure}
\begin{center}

\plotone{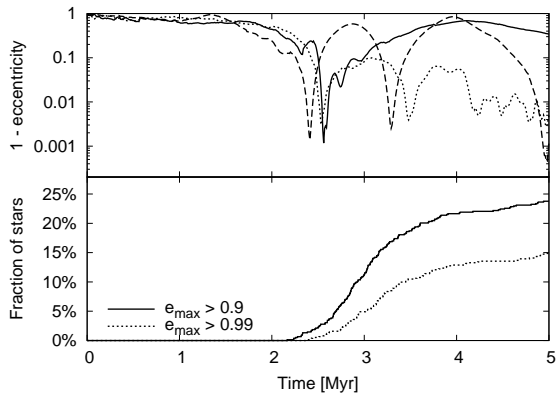}
\caption{
\label{fig:eccent}
Eccentricity evolution of
stellar orbits.
The top panel shows as an example the time evolution of orbital eccentricity of three stars with short Kozai periods.
The cyclic nature of the Kozai mechanism, leading to extremely high eccentricities, is evident.
The solid and dotted lines in the bottom panel depict the fraction of stars having reached eccentricities higher than 0.9 and 0.99, respectively, as a function of time.
Any star on a Kozai oscillation keeps a high eccentricity for a short time only, but as most stars have a Kozai period longer than the computed time,
these numbers are increasing until the end of the integration.
}
\end{center}
\end{figure}

Figure \ref{fig:eccent} shows the eccentricity evolution of some stars in our simulation, and the total fraction of stars
that have reached eccentric orbits within a given time.
A significant fraction of the stars obtain very high eccentricities.
In contrast, a single stellar disk around a SMBH evolves only very slowly: In a corresponding integration,
no star obtained an orbital eccentricity above 0.8 within a comparable time span. This is mainly due to the lack of an external torque,
and the lack of large inclinations required for the Kozai mechanism to become effective.

\section{Disruption of stellar binaries}
\label{sec:disrupt}

As most stars form in binaries \citep{gkgb07},
it is natural to assume that a significant fraction of stars in the stellar disks are followed by a companion.
The orbits of these binaries around the SMBH are subject to the same Kozai mechanism as single stars, but a very close pericenter passage
may break up the binary and leave one star on a tightly bound orbit around the SMBH.

In a set of three-body integrations of a stellar binary orbiting a SMBH, we have investigated the distribution of binary disruption remnants.
As can be seen in Figure \ref{fig:eccent}, the eccentricity evolution due to the Kozai effect is smooth.

\begin{figure}
\plotone{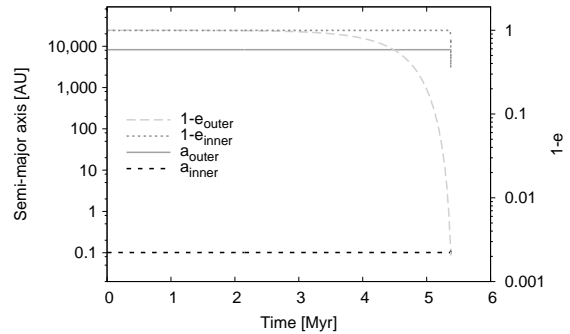}
\caption{
Evolution of orbital parameters of a stellar binary undergoing Kozai resonance. The long-dashed and dotted lines show the outer and inner eccentricity of the binary, respectively, while the solid and short-dashed lines show the outer and inner semimajor axis, respectively.
It is seen that the binary's inner orbital parameters stay constant until binary disruption, while the outer eccentricity grows due to the Kozai mechanism.
\label{fig:binkozai}
}
\end{figure}

We calculated a set of models including a SMBH, an intermediate-mass black hole (IMBH) to drive the Kozai, and a stellar binary on an initially circular orbit at high inclination with respect to the IMBH's orbit in order to test the impact of a Kozai oscillation on the binary's inner orbital parameters.
Figure \ref{fig:binkozai} shows the time evolution of a 0.1\,AU binary's both outer and inner eccentricity and semimajor axis. It can be seen that until binary disruption, only the outer eccentricity is changing.
In particular, the binary does not widen.
Only during the last few pericenter passages does the inner eccentricity grow to a value of $e=0.6$, but this does not significantly change the resulting semimajor axis of the bound member, nor the velocity of the ejected star.
Computations with wider binaries (1\,AU) yield similar results.

\begin{figure}
\plotone{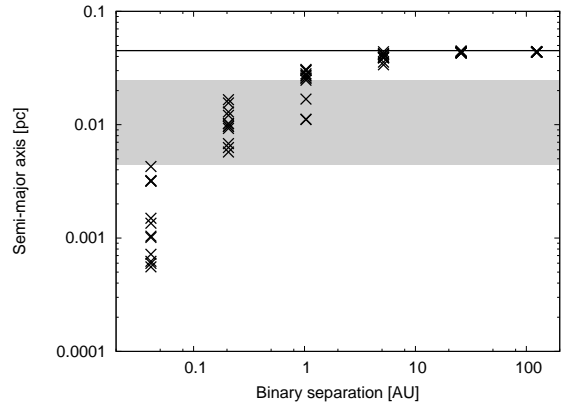}
\caption{
\label{fig:disrupt}
Orbital parameters about the SMBH of the (more strongly) bound member star of a disrupted equal-mass binary as a function of initial binary separation.
While the disruption of wide binaries leaves the individual stars on orbits similar to the initial trajectory,
the bound star of a disrupted tight binary ends up on a very tight orbit around the SMBH.
The solid line depicts the initial semimajor axis before disruption.
The shaded area indicates the semimajor axes of the S stars in the Galactic center.
}
\end{figure}

We thus conclude that any tight binary undergoing Kozai resonance reaches its point of disruption basically unperturbed.
We have therefore computed a set of models where
the pericenter distance of a binary orbiting a SMBH is equal to the disruption distance (Fig.\ \ref{fig:disrupt}).
We find that the disruption remnants of binaries with separations larger than a few AU
orbit the central SMBH on similar trajectories as the initial binary, and so can still be observed as disk stars.
However, the disruption of a tight binary (with separation of 1\,AU or less) forces one star onto a very tight and eccentric S-star-type orbit, while the other star gains a significant amount of energy and may even be ejected from the Galaxy, thus becoming a progenitor of the hypervelocity stars observed in the Galactic halo \citep[e.g.,][]{b07a,gps05}.
For example, disruption of binaries with a separation of 0.2\,AU leaves one star on an orbit with semimajor axis consistent with those of the observed S stars, while the other star is ejected from the Galaxy with a velocity of up to 1000\,km\,s$^{-1}$ in more than half of our integrations.

Binary disruption has been shown to be an effective mechanism to create the S stars \citep{gq03}.
One way to create young close-passage binaries is by massive perturbers like star clusters or molecular clouds at distances of a few pc from the Galactic center \citep{pha06}.
Our simulations show that these binaries are also a natural consequence of the interaction of the two stellar disks:
Using a binary fraction of 30\%, 18 of the B-type binaries achieve pericenter distances below their tidal radius, assuming a binary separation of 0.1\,AU.
Given that not much is known about the binary fraction in the Galactic center, this is fully consistent with the observed number of 15 S stars.

\section{Discussion}
\label{sec:discuss}

The mechanism described above naturally explains the reported mysteries of the S stars:
The disruption of stellar binaries as tight as 0.1--1\,AU can leave a single star orbiting the central SMBH
a factor of 10 closer than the original binary. This is just the distance ratio of S stars and stellar disks observed
in the center of our Galaxy.

The orientation of the disks rapidly changes due to precession. Furthermore, the change of orientation of a binary's orbit
is a natural consequence of the Kozai effect. Once a binary is disrupted and leaves a single star on a much closer orbit, this orbit is
decoupled from the originating stellar disk's motion, as both the precession frequency and the Kozai oscillation period are functions of central distance,
explaining the widely varying orbital orientations observed.

The eccentricities created for the tightly bound stars after tidal disruption are as high as 0.99. Resonant relaxation \citep{rt96} and Kozai interactions
between the S stars are expected to turn these extremely high eccentricities into an isotropic distribution on a short timescale,
just as is observed for the S stars.

While all of the S stars observed so far are main-sequence B-type stars, observations of the stellar disks in the Galactic center
mostly revealed OB giants and Wolf-Rayet stars. But even for the unusually flat mass function suggested \citep{pau06},
one would expect a factor of 5 more B-type main-sequence stars in the disks than O stars. Most of these stars have not been observed yet,
mainly due to crowding of the observed region, limited spatial resolution, and strong nebular emission \citep{pau06}.
Our computation assuming equal-mass binaries predicts only one S star above 20\,$\msun$ per 18 B-type S stars. The lack of observed O-type S stars is therefore not statistically significant.
The number of expected massive S stars would drop further for random pairing of binary component masses, since disruption of a high mass ratio binary will only moderately change the orbit of the massive member.
Furthermore, the massive O stars might not form in as tight binaries as are required to produce S stars.
Hence, it is legitimate to assume that both S stars and disk stars formed in the same environment,
thus providing a solution to the ``paradox of youth''.

In this Letter, we studied the evolution of two isolated stellar disks orbiting a SMBH. In a forthcoming publication, we will analyze the effects of other contributions of a perturbing potential, such as a stellar cusp \citep[e.g.,][]{sea+07}, a massive circumnuclear disk \citep{cssy05}, an intermediate-mass black hole and/or a star cluster \citep[probably IRS 13;][]{mpsr04}.
Preliminary results indicate, e.g., that the presence of a cusp of stellar black holes \citep[e.g.,][]{fak06} does not prevent the dynamical formation of S stars as described in this Letter.

\acknowledgments

We thank Ladislav {\v S}ubr for helpful discussions. This work was supported by the German Research Foundation (DFG) through the priority program 1177 ``Witnesses of Cosmic History: Formation and Evolution of Black Holes, Galaxies and Their Environment.''

\bibliographystyle{apj}

\begin{thebibliography}{}
\expandafter\ifx\csname natexlab\endcsname\relax\def\natexlab#1{#1}\fi

\bibitem[{{Alexander}(2005)}]{a05}
{Alexander}, T. 2005, PhR, 419, 65

\bibitem[{{Alexander} \& {Livio}(2004)}]{al04}
{Alexander}, T., \& {Livio}, M. 2004, \apjl, 606, L21

\bibitem[{{Brown} {et~al.}(2007){Brown}, {Geller}, {Kenyon}, {Kurtz}, \&
  {Bromley}}]{b07a}
{Brown}, W.~R., {Geller}, M.~J., {Kenyon}, S.~J., {Kurtz}, M.~J., \& {Bromley},
  B.~C. 2007, \apj, 660, 311

\bibitem[{{Christopher} {et~al.}(2005){Christopher}, {Scoville}, {Stolovy}, \&
  {Yun}}]{cssy05}
{Christopher}, M.~H., {Scoville}, N.~Z., {Stolovy}, S.~R., \& {Yun}, M.~S.
  2005, \apj, 622, 346

\bibitem[{{Cuadra} {et~al.}(2008){Cuadra}, {Armitage}, \& {Alexander}}]{caa08}
{Cuadra}, J., {Armitage}, P.~J., \& {Alexander}, R.~D. 2008, \mnras, 388, L64

\bibitem[{{Eckart} \& {Genzel}(1997)}]{eg97}
{Eckart}, A., \& {Genzel}, R. 1997, \mnras, 284, 576

\bibitem[{{Eisenhauer} {et~al.}(2005){Eisenhauer}, {Genzel}, {Alexander},
  {Abuter}, {Paumard}, {Ott}, {Gilbert}, {Gillessen}, {Horrobin}, {Trippe},
  {Bonnet}, {Dumas}, {Hubin}, {Kaufer}, {Kissler-Patig}, {Monnet},
  {Str{\"o}bele}, {Szeifert}, {Eckart}, {Sch{\"o}del}, \& {Zucker}}]{eis05}
{Eisenhauer}, F., {Genzel}, R., {Alexander}, T., {Abuter}, R., {Paumard}, T.,
  {Ott}, T., {Gilbert}, A., {Gillessen}, S., {Horrobin}, M., {Trippe}, S.,
  {Bonnet}, H., {Dumas}, C., {Hubin}, N., {Kaufer}, A., {Kissler-Patig}, M.,
  {Monnet}, G., {Str{\"o}bele}, S., {Szeifert}, T., {Eckart}, A.,
  {Sch{\"o}del}, R., \& {Zucker}, S. 2005, \apj, 628, 246

\bibitem[{{Freitag} {et~al.}(2006){Freitag}, {Amaro-Seoane}, \&
  {Kalogera}}]{fak06}
{Freitag}, M., {Amaro-Seoane}, P., \& {Kalogera}, V. 2006, \apj, 649, 91

\bibitem[{{Fuchs} {et~al.}(2006){Fuchs}, {Breitschwerdt}, {de Avillez},
  {Dettbarn}, \& {Flynn}}]{fba06}
{Fuchs}, B., {Breitschwerdt}, D., {de Avillez}, M.~A., {Dettbarn}, C., \&
  {Flynn}, C. 2006, \mnras, 373, 993

\bibitem[{{Genzel} {et~al.}(2003){Genzel}, {Sch{\"o}del}, {Ott}, {Eisenhauer},
  {Hofmann}, {Lehnert}, {Eckart}, {Alexander}, {Sternberg}, {Lenzen},
  {Cl{\'e}net}, {Lacombe}, {Rouan}, {Renzini}, \& {Tacconi-Garman}}]{gs03}
{Genzel}, R., {Sch{\"o}del}, R., {Ott}, T., {Eisenhauer}, F., {Hofmann}, R.,
  {Lehnert}, M., {Eckart}, A., {Alexander}, T., {Sternberg}, A., {Lenzen}, R.,
  {Cl{\'e}net}, Y., {Lacombe}, F., {Rouan}, D., {Renzini}, A., \&
  {Tacconi-Garman}, L.~E. 2003, \apj, 594, 812

\bibitem[{{Gerhard}(2001)}]{ger01}
{Gerhard}, O. 2001, \apjl, 546, L39

\bibitem[{{Ghez} {et~al.}(2005){Ghez}, {Salim}, {Hornstein}, {Tanner}, {Lu},
  {Morris}, {Becklin}, \& {Duch{\^e}ne}}]{ghe05}
{Ghez}, A.~M., {Salim}, S., {Hornstein}, S.~D., {Tanner}, A., {Lu}, J.~R.,
  {Morris}, M., {Becklin}, E.~E., \& {Duch{\^e}ne}, G. 2005, \apj, 620, 744

\bibitem[{{Goodwin} {et~al.}(2007){Goodwin}, {Kroupa}, {Goodman}, \&
  {Burkert}}]{gkgb07}
{Goodwin}, S.~P., {Kroupa}, P., {Goodman}, A., \& {Burkert}, A. 2007, in
  Protostars and Planets V, ed. B.~{Reipurth}, D.~{Jewitt}, \& K.~{Keil},
  133--147

\bibitem[{{Gould} \& {Quillen}(2003)}]{gq03}
{Gould}, A., \& {Quillen}, A.~C. 2003, \apj, 592, 935

\bibitem[{{Gualandris} {et~al.}(2005){Gualandris}, {Portegies Zwart}, \&
  {Sipior}}]{gps05}
{Gualandris}, A., {Portegies Zwart}, S., \& {Sipior}, M.~S. 2005, \mnras, 363,
  223

\bibitem[{{G{\"u}rkan} \& {Hopman}(2007)}]{gh07}
{G{\"u}rkan}, M.~A., \& {Hopman}, C. 2007, \mnras, 379, 1083

\bibitem[{{Hansen} \& {Milosavljevi{\'c}}(2003)}]{hm03}
{Hansen}, B.~M.~S., \& {Milosavljevi{\'c}}, M. 2003, \apjl, 593, L77

\bibitem[{{Holman} {et~al.}(1997){Holman}, {Touma}, \& {Tremaine}}]{htt97}
{Holman}, M., {Touma}, J., \& {Tremaine}, S. 1997, \nat, 386, 254

\bibitem[{{Hurley} {et~al.}(2000){Hurley}, {Pols}, \& {Tout}}]{hpt00}
{Hurley}, J.~R., {Pols}, O.~R., \& {Tout}, C.~A. 2000, \mnras, 315, 543

\bibitem[{{Kozai}(1962)}]{k62}
{Kozai}, Y. 1962, \aj, 67, 591

\bibitem[{{Levin} \& {Beloborodov}(2003)}]{lb03}
{Levin}, Y., \& {Beloborodov}, A.~M. 2003, \apjl, 590, L33

\bibitem[{{L{\"o}ckmann} \& {Baumgardt}(2008)}]{lb08}
{L{\"o}ckmann}, U., \& {Baumgardt}, H. 2008, \mnras, 384, 323

\bibitem[{{Lu} {et~al.}(2006){Lu}, {Ghez}, {Hornstein}, {Morris}, {Matthews},
  {Thompson}, \& {Becklin}}]{lu06}
{Lu}, J.~R., {Ghez}, A.~M., {Hornstein}, S.~D., {Morris}, M., {Matthews}, K.,
  {Thompson}, D.~J., \& {Becklin}, E.~E. 2006, Journal of Physics Conference
  Series, 54, 279

\bibitem[{{Maillard} {et~al.}(2004){Maillard}, {Paumard}, {Stolovy}, \&
  {Rigaut}}]{mpsr04}
{Maillard}, J.~P., {Paumard}, T., {Stolovy}, S.~R., \& {Rigaut}, F. 2004, \aap,
  423, 155

\bibitem[{{Martins} {et~al.}(2008){Martins}, {Gillessen}, {Eisenhauer},
  {Genzel}, {Ott}, \& {Trippe}}]{mge08}
{Martins}, F., {Gillessen}, S., {Eisenhauer}, F., {Genzel}, R., {Ott}, T., \&
  {Trippe}, S. 2008, \apjl, 672, L119

\bibitem[{{McMillan} \& {Portegies Zwart}(2003)}]{mp03}
{McMillan}, S.~L.~W., \& {Portegies Zwart}, S.~F. 2003, \apj, 596, 314

\bibitem[{{Nayakshin}(2005)}]{nay05}
{Nayakshin}, S. 2005, \mnras, 359, 545

\bibitem[{{Nayakshin} \& {Cuadra}(2005)}]{nc05}
{Nayakshin}, S., \& {Cuadra}, J. 2005, \aap, 437, 437

\bibitem[{{Paumard} {et~al.}(2006){Paumard}, {Genzel}, {Martins}, {Nayakshin},
  {Beloborodov}, {Levin}, {Trippe}, {Eisenhauer}, {Ott}, {Gillessen}, {Abuter},
  {Cuadra}, {Alexander}, \& {Sternberg}}]{pau06}
{Paumard}, T., {Genzel}, R., {Martins}, F., {Nayakshin}, S., {Beloborodov},
  A.~M., {Levin}, Y., {Trippe}, S., {Eisenhauer}, F., {Ott}, T., {Gillessen},
  S., {Abuter}, R., {Cuadra}, J., {Alexander}, T., \& {Sternberg}, A. 2006,
  \apj, 643, 1011

\bibitem[{{Perets} {et~al.}(2007){Perets}, {Hopman}, \& {Alexander}}]{pha06}
{Perets}, H.~B., {Hopman}, C., \& {Alexander}, T. 2007, \apj, 656, 709

\bibitem[{{Portegies Zwart} {et~al.}(2003){Portegies Zwart}, {McMillan}, \&
  {Gerhard}}]{pmg03}
{Portegies Zwart}, S.~F., {McMillan}, S.~L.~W., \& {Gerhard}, O. 2003, \apj,
  593, 352

\bibitem[{{Rauch} \& {Tremaine}(1996)}]{rt96}
{Rauch}, K.~P., \& {Tremaine}, S. 1996, New Astronomy, 1, 149

\bibitem[{{Sch{\"o}del} {et~al.}(2007){Sch{\"o}del}, {Eckart}, {Alexander},
  {Merritt}, {Genzel}, {Sternberg}, {Meyer}, {Kul}, {Moultaka}, {Ott}, \&
  {Straubmeier}}]{sea+07}
{Sch{\"o}del}, R., {Eckart}, A., {Alexander}, T., {Merritt}, D., {Genzel}, R.,
  {Sternberg}, A., {Meyer}, L., {Kul}, F., {Moultaka}, J., {Ott}, T., \&
  {Straubmeier}, C. 2007, \aap, 469, 125

\bibitem[{{{\v S}ubr} \& {Karas}(2005)}]{sk05}
{{\v S}ubr}, L., \& {Karas}, V. 2005, \aap, 433, 405

\end{thebibliography}

\makeatletter   \renewcommand{\@biblabel}[1]{[#1]}   \makeatother

\clearpage

\end{document}